\documentstyle[prl,aps]{revtex}

\begin{document}
\author{Jian Qi Shen $^{1,}$$^{2}$ \footnote{E-mail address: jqshen@coer.zju.edu.cn}}
\address{$^{1}$  Centre for Optical
and Electromagnetic Research, State Key Laboratory of Modern
Optical Instrumentation, \\Zhejiang University,
Hangzhou Yuquan 310027, P.R. China\\
$^{2}$ Zhejiang Institute of Modern Physics and Department of
Physics, Zhejiang University, Hangzhou 310027, P.R. China}
\date{\today }
\title{An approach to the time-dependent Jaynes-Cummings model \\without the rotating wave approximation}
\maketitle

\begin{abstract}
This report presents an approach to the exact solutions of the
time-dependent Jaynes-Cummings (J-C) model without the rotating
wave approximation (RWA). It is shown that there is a
squeezing-operator unitary transformation for relating the J-C
model without RWA and the one with RWA. Thus by using an
appropriate squeezing unitary transformation, the time-dependent
J-C model without RWA can be transformed into the one with RWA
that has been exactly solved previously, and based on this one can
readily obtain the exact solutions of the time-dependent
Jaynes-Cummings (J-C) model without RWA. The approach presented
here also shows that we can treat these two kinds of
time-dependent J-C models (both with and without RWA) in a unified
way.
\\ \\
PACS number(s): 03.65.Fd, 42.50.Ct, 03.65.Vf
\end{abstract}

The interaction between a two-level atom and a quantized
single-mode electromagnetic field is described by the
Jaynes-Cummings (J-C) model \cite {Jaynes}, which can be applied
to the study of many quantum effects such as the quantum collapses
and revivals of the atomic inversion, photon antibunching,
squeezing of the radiation inversionless light
amplification\cite{Eberly,Alexanian,Wodkiewicz} as well as
electromagnetically induced transparency\cite
{Imamolglu,Wang,Hong}. Recently, problems of time-dependent
quantum systems attract attention of many researchers in various
areas such as quantum optics\cite{Gong,Shenpla}, condensed matter
physics\cite{Taguchi,Falci}, atomic and molecular physics
(including molecular reaction)\cite {Kuppermann,Kuppermann2,Levi}
and gravity theory\cite {Furtado,Shen1} as well. More recently,
Guedes {\it et al.} solved the one-dimensional Schr\"{o}dinger
equation with a time-dependent linear
potential\cite{Guedes,Bekkar}. As far as we are concerned, by
making use of the Lewis-Riesenfeld invariant theory\cite{Lewis},
we have obtained the exact solutions of the time-dependent
supersymmetric multiphoton J-C model with the rotating wave
approximation\cite{Zhu,Sheneuro}.

It is well known that the J-C model without the rotating wave
approximation (RWA) has close relation to the effects of virtual
photon fields. However, to the best of our knowledge, in the
literature, only the {\it time-dependent} J-C model with RWA and
the {\it time-independent} J-C model without RWA have been exactly
solved. It may be believed that the {\it time-dependent} J-C model
without RWA may get less attention than deserves. The Hamiltonian
of the J-C model without RWA has the non-rotating-wave term, which
makes the Hamiltonian without RWA more complicated than that with
RWA. Historically, maybe for this reason, it was not easy for us
to solve such time-dependent model. In the present report, we will
show that there is a relationship (unitary transformation) between
the J-C model without RWA and the one with RWA: specifically, a
squeezing-operator unitary transformation can relate these two
kinds of J-C model, {\it i.e.}, by using the squeezing-operator
unitary transformation $S(\zeta)=\exp
\left(\frac{\zeta^{\ast}}{2}a^{2}-\frac{\zeta}{2}a^{\dagger 2
}\right)$ with the suitable time-dependent parameters $\zeta$ and
$\zeta^{\ast}$, one can transform the time-dependent J-C model
without RWA into the one with RWA. As stated above, the latter
({\it i.e.}, the time-dependent J-C model with RWA) has been
exactly solved in the literature\cite{Zhu,Sheneuro}. Thus on the
basis of the obtained solutions of the time-dependent J-C model
with RWA, the explicit solutions of the time-dependent J-C model
without RWA will be possibly achieved. In what follows we will
proceed to consider this problem in more detail.

The Hamiltonian of the time-dependent two-level J-C model without
RWA is written in the form (in the unit $\hbar=1$)
\begin{equation}
H(t)=\omega(t)a^{\dagger}a+\frac{\omega_{0}(t)}{2}\sigma_{z}+\gamma(t)a^{\dagger}\sigma_{-}+\gamma^{\ast}(t)a\sigma_{+}+\lambda(t)a\sigma_{-}+\lambda^{\ast}(t)a^{\dagger}\sigma_{+},
\label{eq1}
\end{equation}
where $a^{\dagger}$ and $a$ denote the creation and annihilation
operators of the monomode photon fields interacting with the
two-level atoms characterized by the Pauli's third-component
matrix $\sigma_{z}$; the Pauli matrices $\sigma_{\pm}$ represent
the atomic transition operators satisfying the commutation
relations $[\sigma_{z}, \sigma_{\pm}]=\pm 2\sigma_{\pm}$ and
$[\sigma_{+}, \sigma_{-}]=\sigma_{z}$; the expression
$\lambda(t)a\sigma_{-}+\lambda^{\ast}(t)a^{\dagger}\sigma_{+}$ in
the Hamiltonian (\ref{eq1}) stands for the non-rotating-wave term,
which can describe the effects of virtual photon fields. Note that
here all the coefficients ($\omega, \omega_{0},
\gamma,\gamma^{\ast}, \lambda, \lambda^{\ast}$) in the expression
(\ref{eq1}) are time-dependent parameters (functions). The
time-dependent Schr\"{o}dinger equation governing this
time-dependent J-C model without RWA is
\begin{equation}
H(t)|\Psi(t)\rangle=i\frac{\partial}{\partial t}|\Psi(t)\rangle,
\label{eq2}
\end{equation}
which will be exactly solved in this report.

Unfortunately, due to the presence of the non-rotating-wave term
in the Hamiltonian (\ref{eq1}), it is not easy to obtain the
analytical solutions of the time-dependent Schr\"{o}dinger
equation (\ref{eq2}). In order to overcome this difficulty, in
what follows we will employ a time-dependent squeezing-operator
unitary transformation, which may transform the Hamiltonian
(\ref{eq1}) without RWA into the one with RWA (in which the
non-rotating-wave term is absent). Under this squeezing-operator
transformation $S(\zeta)$ (which will be referred to as
``squeezing unitary transformation''), the time-dependent
Schr\"{o}dinger equation (\ref{eq2}) may be rewritten as follows
\begin{equation}
H_{\rm s}(t)|\Psi_{\rm s}(t)\rangle=i\frac{\partial}{\partial
t}|\Psi_{\rm s}(t)\rangle         \label{eq4}
\end{equation}
with $ |\Psi_{\rm s}(t)\rangle=S^{\dagger}(\zeta)|\Psi(t)\rangle,
\   H_{\rm
s}(t)=S^{\dagger}(\zeta)H(t)S(\zeta)-S^{\dagger}(\zeta)i\frac{\partial}{\partial
t}S(\zeta)$, where the applied squeezing operator takes the form
$S(\zeta)=\exp
\left(\frac{\zeta^{\ast}}{2}a^{2}-\frac{\zeta}{2}a^{\dagger 2
}\right)$. Note that $\zeta=\zeta(t)$ is a time-dependent function
which will be determined in the following. With the help of the
Baker-Campbell-Hausdorff formula\cite{Wei} and the Glauber
formula, one can arrive at

\begin{eqnarray}
-S^{\dagger}(\zeta)i\frac{\partial}{\partial
t}S(\zeta)&=&\frac{-i\left(\dot{\zeta}\zeta^{\ast}-\dot{\zeta}^{\ast}\zeta\right)}{8\zeta^{\ast}\zeta}\left(\cosh
\sqrt{4\zeta^{\ast}\zeta}-1\right)\left(a^{\dagger}a+aa^{\dagger}\right)
 \nonumber \\
 &+&\frac{i\left(\dot{\zeta}\zeta^{\ast}-\dot{\zeta}^{\ast}\zeta\right)}{\left(\sqrt{4\zeta^{\ast}\zeta}\right)^{3}}\left(\sinh
 \sqrt{4\zeta^{\ast}\zeta}-\sqrt{4\zeta^{\ast}\zeta}\right)\left(\zeta^{\ast}a^{2}+\zeta a^{\dagger 2
}\right)-i\left(\frac{\dot{\zeta}^{\ast}}{2}a^{2}-\frac{\dot{\zeta}}{2}a^{\dagger
2 }\right)
\end{eqnarray}
and
\begin{eqnarray}
S^{\dagger}(\zeta)H(t)S(\zeta)&=&\omega(t)\left[-\frac{1}{2}+\frac{\cosh
\sqrt{4\zeta^{\ast}\zeta}}{2}\left(a^{\dagger}a+aa^{\dagger}\right)-\frac{\sinh
\sqrt{4\zeta^{\ast}\zeta}}{\sqrt{4\zeta^{\ast}\zeta}}\left(\zeta^{\ast}a^{2}+\zeta
a^{\dagger 2
}\right)\right]+\frac{\omega_{0}(t)}{2}\sigma_{z}                  \nonumber \\
&+&\left[\gamma\cosh \sqrt{\zeta^{\ast}\zeta}-\frac{\lambda\zeta\sinh \sqrt{\zeta^{\ast}\zeta}}{\sqrt{\zeta^{\ast}\zeta}}\right]a^{\dagger}\sigma_{-}+\left[\gamma^{\ast}\cosh \sqrt{\zeta^{\ast}\zeta}-\frac{\lambda^{\ast}\zeta^{\ast}\sinh \sqrt{\zeta^{\ast}\zeta}}{\sqrt{\zeta^{\ast}\zeta}}\right]a\sigma_{+}                \nonumber \\
&+&\left[\lambda\cosh
\sqrt{\zeta^{\ast}\zeta}-\frac{\gamma\zeta^{\ast}\sinh
\sqrt{\zeta^{\ast}\zeta}}{\sqrt{\zeta^{\ast}\zeta}}\right]a\sigma_{-}+\left[\lambda^{\ast}\cosh
\sqrt{\zeta^{\ast}\zeta}-\frac{\gamma^{\ast}\zeta\sinh
\sqrt{\zeta^{\ast}\zeta}}{\sqrt{\zeta^{\ast}\zeta}}\right]a^{\dagger}\sigma_{+},
\end{eqnarray}
where dot denotes the derivative of $\zeta$ and $\zeta^{\ast}$
with respect to time $t$. If we set
\begin{eqnarray}
\Omega &=&\left[\omega \cosh
\sqrt{4\zeta^{\ast}\zeta}-\frac{i\left(\dot{\zeta}\zeta^{\ast}-\dot{\zeta}^{\ast}\zeta\right)}{4\zeta^{\ast}\zeta}\left(\cosh
\sqrt{4\zeta^{\ast}\zeta}-1\right)\right],  \quad {\mathcal
C}=\left[\frac{\omega}{2}-\frac{i\left(\dot{\zeta}\zeta^{\ast}-\dot{\zeta}^{\ast}\zeta\right)}{8\zeta^{\ast}\zeta}\right]\left(\cosh
\sqrt{4\zeta^{\ast}\zeta}-1\right),                \nonumber \\
{\mathcal A} &=& \frac{\zeta^{\ast}\sinh
\sqrt{4\zeta^{\ast}\zeta}}{\sqrt{4\zeta^{\ast}\zeta}}\left[\frac{i\left(\dot{\zeta}\zeta^{\ast}-\dot{\zeta}^{\ast}\zeta\right)}{4\zeta^{\ast}\zeta}-\omega\right]-\frac{i\left(\dot{\zeta}\zeta^{\ast}-\dot{\zeta}^{\ast}\zeta\right)}{4\zeta^{\ast}\zeta}-\frac{i}{2}\dot{\zeta}^{\ast},
\nonumber \\
g&=&\gamma\cosh \sqrt{\zeta^{\ast}\zeta}-\frac{\lambda\zeta\sinh
\sqrt{\zeta^{\ast}\zeta}}{\sqrt{\zeta^{\ast}\zeta}}, \quad \Lambda
= \lambda\cosh
\sqrt{\zeta^{\ast}\zeta}-\frac{\gamma\zeta^{\ast}\sinh
\sqrt{\zeta^{\ast}\zeta}}{\sqrt{\zeta^{\ast}\zeta}},
\end{eqnarray}
then the obtained Hamiltonian via the squeezing unitary
transformation may be simplified into the following one
\begin{equation}
H_{\rm
s}(t)=\Omega(t)a^{\dagger}a+\frac{\omega_{0}(t)}{2}\sigma_{z}+{\mathcal
A}(t)a^{2}+{\mathcal A}^{\ast}(t)a^{\dagger
2}+g(t)a^{\dagger}\sigma_{-}
+g^{\ast}(t)a\sigma_{+}+\Lambda(t)a\sigma_{-}+\Lambda^{\ast}(t)a^{\dagger}\sigma_{+}+{\mathcal
C}(t).    \label{eq3}
\end{equation}
Evidently, if the two constraints, ${\mathcal A}(t)=0$ and
$\Lambda(t)=0$, are satisfied, then the quadratic-form terms
$a^{2}$ and $a^{\dagger 2}$, and the non-rotating-wave term in the
Hamiltonian (\ref{eq3}) will be absent. It should be further
pointed out that the two equations ${\mathcal A}(t)=0$ and $
\Lambda(t)=0$ are just employed to determine the time-dependent
functions $\zeta$ and $ \zeta^{\ast}$ in the squeezing operator
$S(\zeta)$. Namely, only when the time-dependent functions $\zeta$
and $ \zeta^{\ast}$ agree with these two equations will the
squeezing operator $S(\zeta)$ change $H(t)$ without RWA into the
form with RWA (see in the following). Thus we obtain the
time-dependent Hamiltonian without the non-rotating-wave term,
{\it i.e.}, $ H_{\rm
s}(t)=\Omega(t)a^{\dagger}a+\frac{\omega_{0}(t)}{2}\sigma_{z}+g(t)a^{\dagger}\sigma_{-}
+g^{\ast}(t)a\sigma_{+}+{\mathcal C}(t) $. It is readily verified
that if $|\Psi_{\rm s}(t)\rangle$ is rewritten as $|\Psi_{\rm
s}(t)\rangle=\exp [\frac{1}{i}\int^{t}_{0}{\mathcal C}(t'){\rm
d}t']|\Psi_{\rm s}^{\rm RWA}(t)\rangle$, then the time-dependent
Schr\"{o}dinger equation (\ref{eq4}) may be rewritten in the
following form
\begin{equation}
H_{\rm s}^{\rm RWA}(t)|\Psi_{\rm s}^{\rm
RWA}(t)\rangle=i\frac{\partial}{\partial t}|\Psi_{\rm s}^{\rm
RWA}(t)\rangle                   \label{eqq6}
\end{equation}
with $H_{\rm s}^{\rm RWA}(t)=H_{\rm s}(t)-{\mathcal C}(t)$, {\it
i.e.},
\begin{equation}
H_{\rm s}^{\rm
RWA}(t)=\Omega(t)a^{\dagger}a+\frac{\omega_{0}(t)}{2}\sigma_{z}+g(t)a^{\dagger}\sigma_{-}
+g^{\ast}(t)a\sigma_{+},       \label{eq6}
\end{equation}
which is a standard form of the time-dependent Hamiltonian of the
J-C model with RWA. Thus we have shown that there is a connection
(unitary transformation) between the two kinds of time-dependent
J-C model (without RWA and with RWA), and that under the so-called
squeezing unitary transformation, the complicated time-dependent
J-C model without RWA can be reduced to the simple J-C model with
RWA.

To close this report, we will briefly discuss the exact solutions
of the time-dependent J-C model with RWA characterized by
Eq.(\ref{eq6}). It is well known that the Lewis-Riesenfeld
invariant theory\cite{Lewis} can be applied to the exact solutions
of the time-dependent quantum models, the algebraic generators in
the Hamiltonians of which form a certain closed Lie algebra.
Unfortunately, the commutation relations among the generators in
the Hamiltonian (\ref{eq6}) are $[a\sigma_{+},
a^{\dagger}\sigma_{-}]=N'\sigma_{z}$, $[a\sigma_{+},
\sigma_{z}]=-2a\sigma_{+}$,
$[a^{\dagger}\sigma_{-},\sigma_{z}]=2a^{\dagger}\sigma_{-}$, which
shows that $\{a\sigma_{+}, a^{\dagger}\sigma_{-}, \sigma_{z},
N'\}$ do not form a closed Lie algebra. But note that here $N'$ is
of the form
\begin{eqnarray}
N'=\left(
\begin{array}{cc}
aa^{\dagger } & 0 \\
0 & a^{\dagger }a
\end{array}
\right),
\end{eqnarray}
which commutates with all the generators in the Hamiltonian
(\ref{eq6}), namely, $[N', H_{\rm s}^{\rm RWA}(t)]=0$. This,
therefore, means that the operator $N'$ is just a time-independent
Lewis-Riesenfeld invariant that agrees with the Liouville-Von
Neumann equation\cite{Lewis} and that one can obtain a closed
quasi-algebra $\{a\sigma_{+}, a^{\dagger}\sigma_{-}, \sigma_{z}\}$
in the sub-Hilbert-space of $N'$ so long as $N'$ in the
commutation relation $[a\sigma_{+},
a^{\dagger}\sigma_{-}]=N'\sigma_{z}$ is replaced with its certain
eigenvalue\cite{Zhu,Sheneuro}. Further analysis shows that $N'$
possesses the eigenvalue $m+1$ and that the corresponding
eigenvalue equation can be written in the form

 \begin{equation}
N'{%
{\left| m\right\rangle  \choose \left| m+1\right\rangle }%
}=(m+1){%
{\left| m\right\rangle  \choose \left| m+1\right\rangle }%
},                \label{eeq}
\end{equation}
where $\left| m\right\rangle$ and $\left| m+1\right\rangle$ denote
the number states of photon field with the corresponding
occupation numbers being $m$ and $m+1$, respectively.

 Hence, by working in the sub-Hilbert-space of $N'$ corresponding to the
eigenvalue $m+1$, we can exactly solve Eq.(\ref{eqq6}) of the
time-dependent J-C model with RWA by using the Lewis-Risenfeld
invariant theory\cite{Lewis} and the invariant-related unitary
transformation formulation\cite{Gao}. In the following we will
present only the final results without providing the mathematical
procedure. For the detailed calculation, readers may be referred
to Ref.\cite{Sheneuro}.

In the Lewis-Riesenfeld invariant theory\cite{Lewis,Gao}, it is
readily verified that the time-dependent unitary transformation $
V(t)=\exp
\left[\beta(t)a^{\dagger}\sigma_{-}-\beta^{\ast}(t)a\sigma_{+}\right]
$ with
$
\beta(t)=-\frac{\theta(t)\exp [-i\phi(t)]}{2\sqrt{m+1}}, \quad
\beta^{\ast}(t)=-\frac{\theta(t)\exp [i\phi(t)]}{2\sqrt{m+1}}
$
will transform the {\it time-dependent} Lewis-Riesenfeld
invariant\cite{Lewis,Sheneuro}
\begin{equation}
{\mathcal I}(t)=-\frac{\sin \theta(t)}{\sqrt{m+1}}\left[\exp
[-i\phi(t)]a^{\dagger}\sigma_{-}+\exp [i\phi(t)]a\sigma_{+}
\right]+\cos \theta(t)\sigma_{z}                \label{eq10}
\end{equation}
into a {\it time-independent} one, {\it i.e.},
$V^{\dagger}(t){\mathcal I}(t)V(t)=\sigma_{z}$\cite{Sheneuro}. The
eigenvalue of $\sigma_{z}$ is $\sigma=\pm 1$. So, in accordance
with Eq.(\ref{eeq}), in the sub-Hilbert-space of $N'$ the two
eigenstates of the Lewis-Riesenfeld invariant (\ref{eq10})
corresponding to the eigenvalues $\sigma=\pm 1$ are $V(t){\left|
m\right\rangle \choose 0 }$, $V(t){0 \choose \left|
m+1\right\rangle }$, respectively. According to the
Lewis-Riesenfeld theory\cite{Lewis}, the solutions of the
time-dependent Schr\"{o}dinger equation (\ref{eqq6}) can be
constructed in terms of the eigenstates of the invariant
(\ref{eq10}), one can therefore obtain the following two explicit
solutions ({\it i.e.}, without chronological product)
\begin{eqnarray}
|\Psi_{\rm s}^{\rm RWA}(\sigma=+1, N'=m+1; t)\rangle &=&\exp
\left[\frac{1}{i}\int^{t}_{0}\dot{\varphi}_{\sigma=+1}(t'){\rm
d}t'\right]V(t){\left| m\right\rangle  \choose 0 },
                  \nonumber \\
|\Psi_{\rm s}^{\rm RWA}(\sigma=-1, N'=m+1; t)\rangle &=&\exp
\left[\frac{1}{i}\int^{t}_{0}\dot{\varphi}_{\sigma=-1}(t'){\rm
d}t'\right]V(t){0 \choose \left| m+1\right\rangle }, \label{eq20}
\end{eqnarray}
of Eq.(\ref{eqq6}) by working in the sub-Hilbert-space of $N'$
corresponding to the eigenvalue $m+1$, where the integrand
$\dot{\varphi}_{\sigma}(t)$ is defined to be\cite{Sheneuro}
\begin{equation}
\dot{\varphi}_{\sigma}(t)=\left(m+\frac{1}{2}\right)\Omega(t)-\frac{1}{2}\sigma\left\{\sqrt{m+1}\left[g(t)\exp
[i\phi(t)]+g^{\ast}(t)\exp [-i\phi(t)]\right]\sin \theta
(t)+[\Omega(t)-\omega_{0}(t)]\cos \theta(t)\right\}.
\end{equation}
Hence, by the aid of the solutions of Eq.(\ref{eq4}) and
(\ref{eqq6}), one can finally obtain the exact solutions of the
time-dependent Schr\"{o}dinger equation (\ref{eq2}) without RWA,
which are as follows
\begin{eqnarray}
|\Psi(\sigma=+1, N'=m+1; t)\rangle &=&\exp
\left\{\frac{1}{i}\int^{t}_{0}\left[{\mathcal C
}(t')+\dot{\varphi}_{\sigma=+1}(t')\right]{\rm
d}t'\right\}S(\zeta)V(t){\left| m\right\rangle  \choose 0 },                 \nonumber \\
|\Psi(\sigma=-1, N'=m+1; t)\rangle &=&\exp
\left\{\frac{1}{i}\int^{t}_{0}\left[{\mathcal C
}(t')+\dot{\varphi}_{\sigma=-1}(t')\right]{\rm
d}t'\right\}S(\zeta)V(t){0 \choose \left| m+1\right\rangle }.
\end{eqnarray}

To summarize, we have shown that the time-dependent J-C model
without RWA can be transformed into the one with RWA under a
squeezing unitary transformation, which enables us to easily
obtain the exact solutions of the time-dependent J-C model without
RWA. Moreover, in this report the fact that by using the squeezing
unitary transformation one can treat the two time-dependent J-C
models (both with and without RWA) in a unified way is
demonstrated, which also implies that we can consider the quantum
effects of virtual photons in the J-C model from the point of view
of the squeezing transformation. In addition, we think that the
method presented here may seem to be somewhat ingenious and
therefore deserves further consideration in its application to
other quantum-mechanical models without RWA.

\textbf{Acknowledgements}  The author is deeply indebted to X.C.
Gao for many helpful suggestions and enlightening discussions
regarding the topics on the sub-Hilbert-space of the
time-independent Lewis-Riesenfeld invariant. This work was
supported partially by the National Natural Science Foundation of
China under Project No. $90101024$.

\end{document}